\documentclass{iaus}
\usepackage{graphicx}

\title[The Star Formation History of the GRB 050730 Host Galaxy] 
{The Star Formation History of the GRB 050730 Host Galaxy}

\author[Francesco Calura]   
{Francesco Calura$^1$}

\affiliation{$^1$ Dipartimento di Astronomia, 
Universita' di Trieste, via G. B. Tiepolo 11, 34131 TRIESTE - ITALY
\\email: {\tt fcalura@oats.inaf.it}}

\pubyear{2008}
\volume{255}  
\pagerange{119--126}
\jname{Low-Metallicity Star Formation: From the First Stars to Dwarf Galaxies}
\editors{L.K. Hunt, S. Madden \& R. Schneider, eds.}
\begin{document}

\maketitle

\begin{abstract}
The long GRB 050730 observed at redshift $z \sim 4$ allowed the determination 
of the elemental abundances for a set of different chemical elements. 
We use detailed chemical evolution models taking into account also dust production 
to constrain the star formation history of the host galaxy of this long GRB. 
For the host galaxy of GRB 050730, 
we derive also some dust-related quantities and the the specific star formation rate, namely the 
star formation rate per unit stellar mass.  
We copare the properties of the GRB host galaxy with the ones 
of Quasar  Damped Lyman Alpha absorbers. 
\end{abstract}

\firstsection 
\section{Introduction}
Gamma-ray bursts (GRBs) afterglows provide an insight 
into the interstellar medium (ISM) of galaxies 
during the earliest stages of their evolution. 
In several cases, thanks to the GRB afterglows it has been possible 
to study the dust content, the star formation rates and the stellar mass 
of the GRB host (Bloom et al. 1998; Savaglio et al. 2008). \\
In a few cases, 
the determination of their chemical abundance pattern has been possible 
(Savaglio et al. 2003; Vreeswijk et al. (2004); Prochaska et al. 2007). \\
Chen et al. (2005)  reported on the
chemical abundances for the damped Lyman Alpha system (DLA) associated
with the host galaxy of GRB 050730.  
Their analysis showed that this gas was metal poor with
modest depletion. 
These results were subsequently
expanded and tabulated by Prochaska et  al. (2007) (hereafter P07). \\
 In this paper, we aim at determining the star formation history 
of a GRB host galaxy by studying  the chemical abundances measured 
in the afterglow spectrum of GRB 050730. For this purpose, we use a detailed chemical evolution 
model. 
Our aim is to constrain  
the star formation 
rate and the age of the host galaxy of GRB 050730, and possibly to expand our study to other systems.\\
The plan of this paper is as follows. 
In Section 2, we briefly introduce the chemical evolution model. 
In Section 3, we present and discuss our main results.

\begin{figure}[b]
\begin{center}
 \includegraphics[width=3.4in]{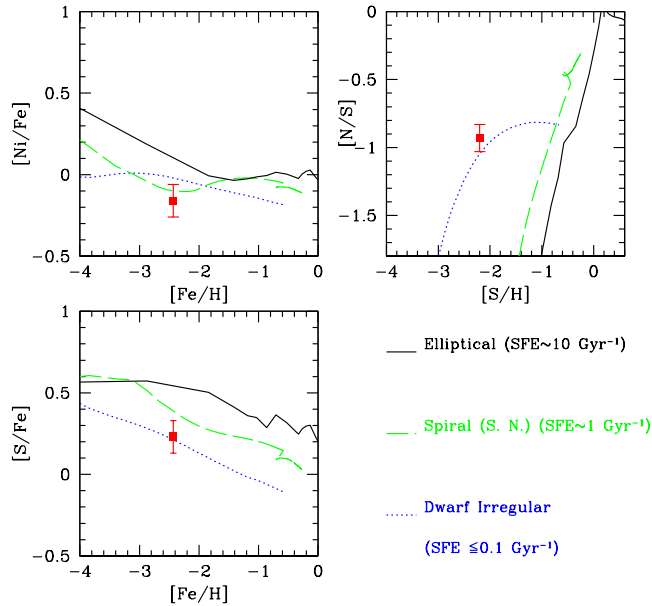} 
 \caption{Observed abundance ratios versus metallicity for the host galaxy of GRB 050730   
as reported by P07 (solid squares with error bars). 
The solid line, dashed line and dotted line 
are predictions computed by means of a chemical evolution model for an elliptical galaxy, a spiral galaxy and a dwarf 
irregular galaxy, respectively, and without including dust depletion (see text for further details).}
   \label{fig1}
\end{center}
\end{figure}

\section{Chemical evolution models including dust}

The chemical evolution model used here to derive the star formation history of  the GRB 050730 host galaxy is similar to that 
developed by Bradamante et al. (1998) for  dwarf irregular galaxies.  
Since some chemical species studied in this paper are refractory (Fe, Ni),  
in the chemical evolution model  
we include also a treatment of dust production and destruction, based on  the work by Calura, Pipino \& Matteucci (2008, hereinafter CPM08). \\
We use the model for dwarf irregular galaxies since several 
observational investigations  
provided strong evidence that most of the GRBs originate 
in gas rich, star-forming sub-luminous ($L<L_{*}$) galaxies with relatively low 
metallicities ($Z < Z_{\odot}$) (Bloom et al. 1998; Prochaska et al. 2004). \\
The dwarf galaxy is assumed to form by means 
of a continuous infall of 
pristine gas until a mass $M_{tot}$ is accumulated.  
The evolution of dwarf irregular galaxies is characterized  
by a continuous star formation history, characterized by a low star formation efficiency 
($\nu \le 0.1 Gyr^{-1}$). \\
The star 
formation rate $\psi(t)$ is directly proportional the gas 
fraction $G(t)$ at the time $t$, according to the Schmidt law: 
\begin{equation}
\psi(t)=\nu G(t).
\end{equation}
Supernovae (SNe) Ia and II are responsible for the onset of 
a galactic wind, when the thermal 
energy of the ISM exceeds its binding energy, 
which is related to the presence of a dark matter 
halo in which the galaxy is embedded (for more details, see Bradamante et al. 1998, Lanfranchi \& Matteucci 2003). 
The binding energy of the gas is 
influenced by  assumptions concerning the presence and distribution
of dark matter (Matteucci 1992). A diffuse ($R_e/R_d$=0.1,  where
$R_e$ is the effective radius of the galaxy and $R_d$ is  the radius
of the dark matter core) and massive  ($M_{dark}/M_{Lum}=10$) dark
halo has been assumed for each galaxy. 
The time at which the wind develops depends on the assumed star formation efficiency. 
In general, the higher the star formation (SF) efficiency, the earlier the wind develops. 
The models used throughout this paper is characterized by very low star formation efficiencies 
and by a young age, much lower than the times of occurrence of the galactic winds. 
For these reasons, the galactic winds have no effect on the main results obtained in this paper. \\
Chemical enrichment from various types of stars is properly taken into account. 
The stellar yields are mainly from Woosley \& Weaver (1995) for 
massive stars, from Meynet \& Maeder (2002) for low and intermediate mass stars (LIMS) 
and from Iwamoto et al. (1999) 
for Type Ia SNe. \\
We assume a Salpeter initial mass function (IMF). 
We assume a cosmological model characterized by $\Omega_m=0.3$, 
$\Omega_{\Lambda}=0.7$ and a Hubble constant $H_{0}=70 km s^{-1} Mpc^{-1}$.

\subsection{Dust production and destruction} 
Two  elements studied in these paper are refractory:  Ni and Fe. 
To model their gas phase 
abundances, we need to take into account dust production and desctruction.  \\
The model for dust evolution used in this paper is described in detail in   CPM08. 
Here, we summarize its  main features. 
For the refractory chemical element labeled $i$, 
a fraction $\delta^{SW}_{i}$, $\delta^{Ia}_{i}$, and  $\delta^{II}_{i}$ is incorporated in dust grains by 
low and intermediate mass stars, type Ia SNe, and type II SNe, respectively. These quantities are the dust condensation efficiencies 
of the element $i$ in various stellar objects. 
Here we assume $\delta^{SW}_{i}=\delta^{Ia}_{i}=\delta^{II}_{i} \equiv \delta_{i} =0.1$. 
This choice is motivated by recent mid-infrared observations of one Supernova 
(Sugerman et al. 2003), which provided an upper limit of $\delta^{II}_{i} \le 0.12$. 
The value assumed here is further supported by  theoretical studies of the local dust cycle 
(Edmunds 2001)\\ 
Dust grains are usually  destroyed by the propagation 
of SN shock waves in the warm/ionised interstellar medium (Jones et al. 1994). 
If $G_{dust,i}$ is the fraction of the element $i$ locked into dust and $G$ is the gas fraction, 
the destruction rate is calculated as 
$G_{dust,i}/\tau_{destr}$, where $\tau_{destr}$ is the dust destruction timescale, calculated as: 
\begin{equation}
\tau_{destr}=(\epsilon M_{SNR})^{-1} \frac{G}{R_{SN}}. 
\end{equation} 
(McKee 1989). 
$M_{SNR}=1300 M_{\odot}$ is the mass of the interstellar gas swept up by the SN remnant. 
$R_{SN}$ is the total SNe rate, i.e. the sum of the rates of Type Ia and Type II SNe.\\ 
Unless otherwise specified, 
we assume that no dust accretion, occurring mainly in cold molecular clouds,  is taking place in the GRB host galaxy. 
Our choice is motivated by the fact that 
very little molecular H is observed in local dwarfs, with molecular-to-atomic gas fractions 
of $\sim 10 \%$ or lower (Clayton et al. 1996). Our assumption is  
further supported by the lack of $H_{2}$ absorption lines observed in the spectra of 
GRB afterglows (Whalen et al. 2008). 
In the presence of intense SF, molecular clouds are likely to rapidly dissolve, 
allowing very little dust accretion to occurr.\\
\begin{figure}
\includegraphics[width=\columnwidth]{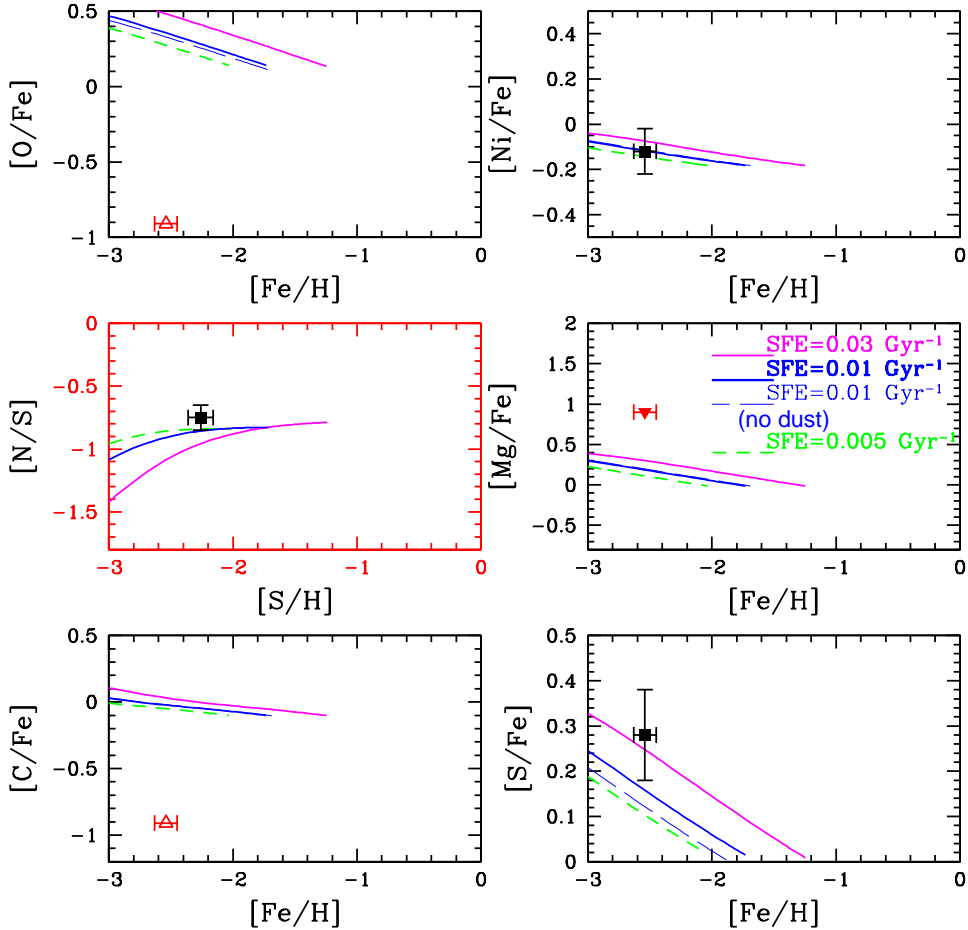}
\caption{ Observed abundance ratios versus metallicity for the host galaxy of GRB 050730   
as derived by P07 (solid squares with error bars).
The triangles stand for upper or lower limits.  
The thick lines represent models with increasing SF efficiency, 
with the lowest curves having the lowest SF efficiency values. 
The thin dashed lines do not include dust depletion. The red box (N/S vs S/H) 
is used for abundance ratios between  non-refractory elements.  }
\label{fig2}
\end{figure}
\begin{figure}
\includegraphics[width=\columnwidth]{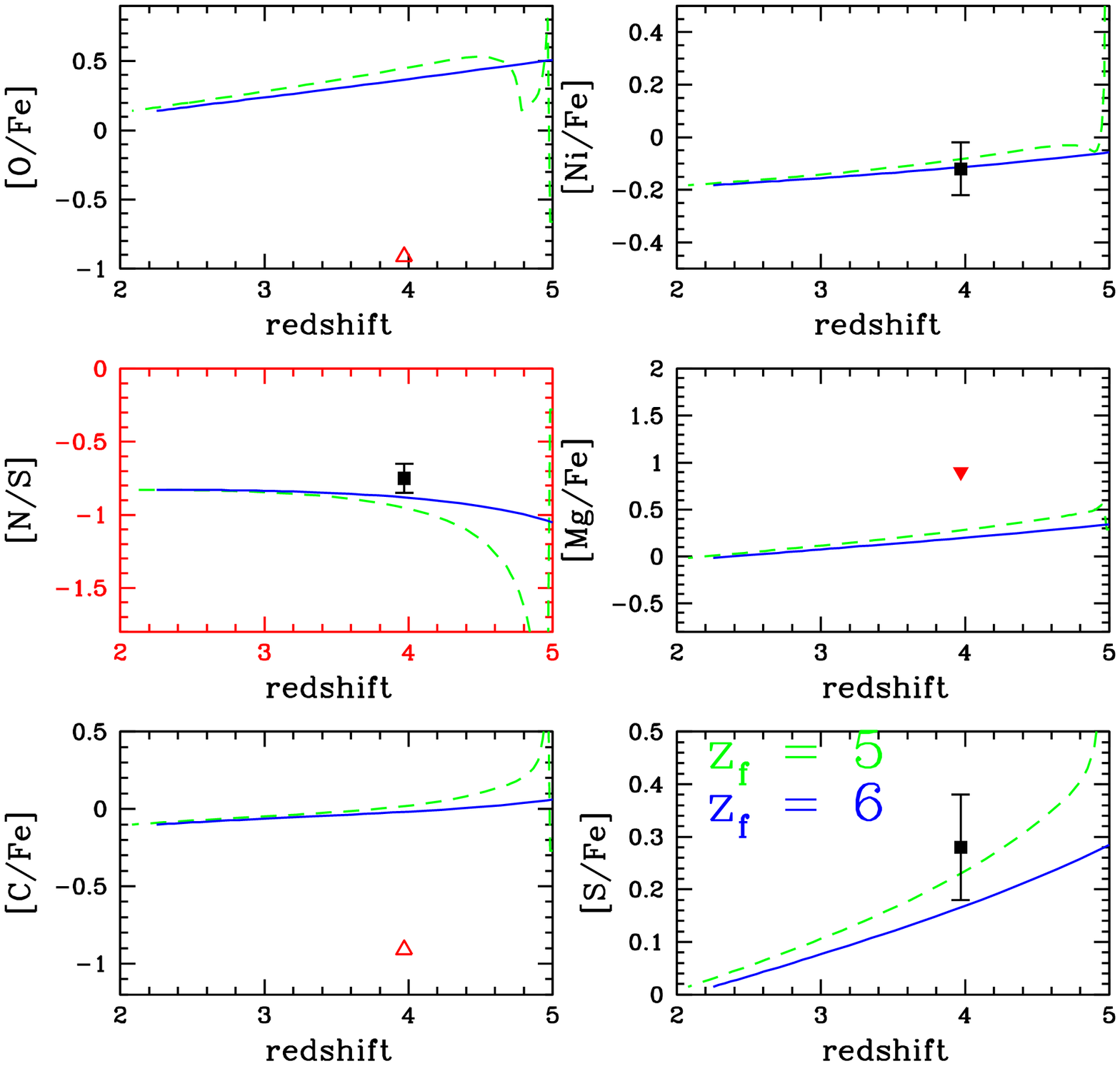}
\caption{ Observed abundance ratios versus redshift for the host galaxy of GRB 050730   
as derived by P07 (solid squares with error bars). 
The triangles stand for upper or lower limits. 
The solid (dashed) lines represent a redshift of formation $z_f=6$  
($z_f=5$). 
The thin dashed lines do not include dust depletion. The red box (N/S vs redshift) 
is used for abundance ratios between  non-refractory elements.  }
\label{fig3}
\end{figure}

\subsection{Results and discussion}
The abundance ratios between two 
elements formed on 
different timescales can be used as "cosmic clocks" and provide us with information on the roles of 
LIMS and SNe in the enrichment process (Matteucci 2001). 
In particular, the study of abundance ratios such 
as [$\alpha$/Fe] and [N/$\alpha$]
\footnote{All the abundances between two different elements X and Y 
are expressed as $[X/Y]=log(X/Y)-log(X/Y)_{\odot}$, where  $(X/Y)$ and $(X/Y)_\odot$ are 
the ratios between the mass fractions of X and Y in the ISM and in the sun, respectively. 
We use the set of solar abundances as determined by Grevesse et al. (2007).} 
is quite helpful, since the $\alpha$-elements (O, S, Mg) 
are produced on short timescales by Type II 
SNe, whereas the Fe-peak elements and nitrogen are produced on lon 
timescales by Type Ia SNe and low and intermediate-mass stars, respectively. 
As shown by Dessauges-Zavadsky at al. (2004; 2007, hereinafter D07), 
the simultaneous study of the abundance ratios between different elements as functions 
of a metallicity tracer (such as [S/H] or [Fe/H] ) may be used  to constrain the star formation history 
of a given system, 
 whereas the study of the abundance ratios versus redshift can be used 
to constrain the age of the system. \\
In Fig. 1, we show the predicted evolution of [N/Fe], [Ni/Fe] and [S/Fe] as a function of 
various metallicity tracers, such as [Fe/H] or [S/H], for various star formation histories 
describing galaxies of different morphological types: an elliptical galaxy (solid lines), 
a spiral galaxy (more precisely, a model for the Solar Neigbourhood, dashed lines, see CPM08 for a detailed model description) and a dwarf irregular galaxy 
(dotted lines), compared with the values observed for GRB 050730 DLA (solid squares with error bars).  
Fig 1 shows  clearly that the most likely progenitor for the host galaxy of GRB 050730 is a dwarf irregular galaxy, 
and how the elliptical and spiral model appear indadequate to describe the abundance pattern observed by P07 in the 
GRB 050730 host. \\
In the following, we will use the dwarf irregular galaxy model to constrain the parameters 
of the SF history and the age of the host galaxy  GRB 050730. \\
The next step is to  constrain 
the main star formation history parameter of  the host galaxy of GRB 050730, i.e. its star formation 
efficiency $\nu$. We search for the model which reproduces at best the 
observed abundances. 
In figure ~\ref{fig2}, we show the predicted evolution of several  abundance ratios 
for various models, characterized by different SF eficiencies.  
The model which reproduces at best the observed 
abundances is the one characterized by a SF efficiency $\nu = 0.01 Gyr^{-1}$.\\
Once we have determined the star formation efficiency of the best model,  
the following step is to constrain the age of the GRB host galaxy. 
To perform this task, we study the abundance ratios vs redshift (Fig.~\ref{fig3}). 
The abundances are reproduces at best at an ages of $\sim 0.4$ Gyr, corresponding 
to a redshift of formation $z_f \sim 5$. 
At this age, the specific star formation rate is $\sim 5 $ Gyr$^{-1}$, 
the Dust-to-Gas ratio is $\sim 10^{-6}$, in agreement with the upper limit of $8 \cdot  10^{-6}$ as derived observationally 
by P07. The Dust-to-metals ratio is 0.04,  consistent with the upper limit of 0.1  measured by P07. \\
\begin{figure}
\includegraphics[width=\columnwidth]{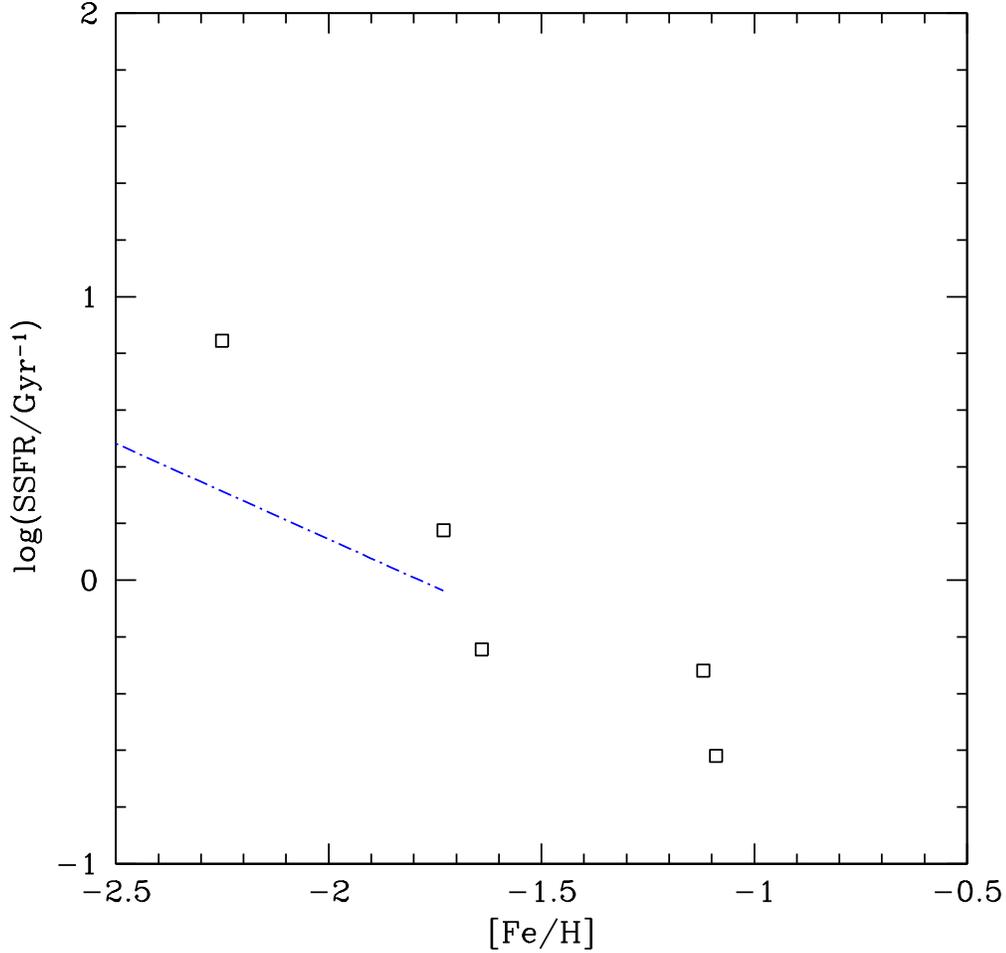}
\caption{Predicted evolution of the SSFR vs [Fe/H] for the best model for the host galaxy of GRB 050730.  
The open squares are the values 
determined for 5 QSO DLAs in the sample by D07. }
\label{fig4}
\end{figure}
Finally, in Fig.~\ref{fig4}, we compare the predicted SSFR vs [Fe/H] obtained for the best model 
for the host galaxy of GRB 050730, with SSFR derived  for a sample of QSO DLAs by D07. 
The evolution of the SSFR of the GRB host is compatible with the values found for QSO DLAs, possibly 
indicating similar chemical evolution paths. However, this may be the result of a coincidence, since in principle 
GRB DLAs are expected to represent denser and more metal-enriched galactic regions than QSO DLAs. 
By extending the method presented here to other systems, it will be possible to shed light on 
possible analogies and differences between QSO and GRB DLAs.

\acknowledgments
It is a pleasure to thank my collaborators, i.e.  Jason Prochaska, Mirka Dessauges-Zavasky and Francesca Matteucci.\\ 
I acknowledge financial contribution from contract ASI-INAF I/016/07/0.\\


\begin{thebibliography}{}
\bibitem[Bloom et al. (1998)]{BLO98} Bloom, J. S.; Djorgovski, S. G.; Kulkarni, S. R.; Frail, D. A., 1998, ApJ, 507, L25
\bibitem[Bradamante et al. (1998)]{BRA98} Bradamante, F., Matteucci, F., D'Ercole, A. 1998, A\&A,{\bf337}, 338
\bibitem[Calura et al. (2008)]{CAL08} Calura, F.; Pipino, A.; Matteucci, F.,  2008, A\&A, 479, 669 (CPM08)
\bibitem[Clayton et al. (1996)]{CLA96} Clayton, Geoffrey C., Green, J., Wolff, Michael J., Zellner, Nicolle E. B., Code, A. D., Davidsen, Arthur F., WUPPE Science Team, HUT Science Team, 1996, ApJ, 460 313 
\bibitem[Dessauges-Zavadsky et al. (2004)]{DES04} Dessauges-Zavadsky, M.; Calura, F.; Prochaska, J. X.; D'Odorico, S.; Matteucci, F., 2004, A\&A, 416, 79
\bibitem[Dessauges-Zavadsky et al. (2007)]{DES07} Dessauges-Zavadsky, M.; Calura, F.; Prochaska, J. X.; D'Odorico, S.; Matteucci, F., 2007, A\&A, 470, 431 (D07)
\bibitem[Edmunds (2001)]{EDM01} Edmunds, M. G., 2001, MNRAS, 328, 223
\bibitem[Grevesse et al. (2007)]{GRE07} Grevesse, N.; Asplund, M.; Sauval, A. J.,  2007, SSRv, 130, 105
\bibitem[Jones et al. (1994)]{JON94} Jones, A. P., Tielens, A. G. G. M., Hollenbach, D. J., McKee, C. F., 1994, ApJ, 433, 797
\bibitem[Iwamoto et al. (1999)]{IWA99} Iwamoto, K.; Brachwitz, F.; Nomoto, K.; Kishimoto, N.; Umeda, H.; Hix, W. R.; Thielemann, F.-K., 1999, ApJS, 125, 439
\bibitem[Lanfranchi \& Matteucci (2003)]{LAN03} Lanfranchi, G., Matteucci, F., 2003, MNRAS, 345, 71 
\bibitem[Matteucci (1992)]{MAT92} Matteucci, F., 1992, ApJ, 397, 32
\bibitem[Matteucci (2001)]{MAT01} Matteucci, F., 2001, \emph{The chemical evolution of the Galaxy}, Astrophysics and space science library, Volume 253, Dordrecht: Kluwer Academic Publishers
\bibitem[McKee (1989)]{MCK89} McKee C. F., 1989, in Allamandola L. J., Tielens A. G. G. M., eds, Interstellar Dust, Proc. IAU Symposium 135. Kluwer, Dordrecht, p. 431
\bibitem[Meynet \& Maeder (2002)]{MEY02} Meynet, G., Maeder, A., 2002, A\&A, 381, L25
\bibitem[Prochaska et al. (2004)]{PRO04}Prochaska, J. X.; Bloom, J. S.; Chen, H.-W.; Hurley, K. C.; Melbourne, J.; Dressler, A.; Graham, J. R.; Osip, D. J.; Vacca, W. D., 2004, ApJ, 611, 200
\bibitem[Prochaska et al. (2007)]{PRO07} Prochaska, J. X.; Chen, H.-W.; Dessauges-Zavadsky, M.; Bloom, J. S., 2007, ApJ, 666, 267
\bibitem[Savaglio et al.(2003)]{SAV03} Savaglio, S., Fall, S. M., Fiore, F.,  2003, ApJ, 585, 638
\bibitem[Savaglio et al.(2008)]{SAV08} Savaglio, S., Glazebrook, K., Le Borgne, D., 2008, ApJ, submitted , arXiv0803.271
\bibitem[Sugerman et al. (2006)]{SUG06} Sugerman, B. E. K., et al., 2006, Science, 313, 196
\bibitem[Vreeswijk et al. (2004)]{VRE04} Vreeswijk, P. M., et al., 2004, A\&A, 419, 927
\bibitem[Whalen et al. (2008)]{WHA08} Whalen, D., Prochaska, J. X., Heger, A., Tumlinson, J., 2008, ApJ, submitted
\bibitem[Woosley \& Weaver (1995)]{WOO95} Woosley, S.E., Weaver, T.A., 1995, ApJS, 101, 181
%


\end{thebibliography}
\end{document}